\documentclass[aps,prl,showpacs,superscriptaddress,twocolumn]{revtex4-1}
\usepackage{graphics} 
\usepackage{amssymb}
\usepackage{dcolumn}
\usepackage{bm} 
\begin{document}
\title{Intrinsic spin-orbit coupling in superconducting $\delta$-doped SrTiO$_3$ heterostructures}
\author{M. Kim}
\affiliation{Department of Advanced Materials Science, University of Tokyo, Kashiwa, Chiba 277-8651, Japan}
\author{Y. Kozuka}
\affiliation{Department of Advanced Materials Science, University of Tokyo, Kashiwa, Chiba 277-8651, Japan}
\author{C. Bell}
\affiliation{Department of Advanced Materials Science, University of Tokyo, Kashiwa, Chiba 277-8651, Japan}
\affiliation{Department of Applied Physics and Stanford Institute for Materials and Energy Science, Stanford University, Stanford, California 94305, USA}
\author{Y. Hikita}
\affiliation{Department of Advanced Materials Science, University of Tokyo, Kashiwa, Chiba 277-8651, Japan}
\author{H. Y. Hwang}
\affiliation{Department of Advanced Materials Science, University of Tokyo, Kashiwa, Chiba 277-8651, Japan}
\affiliation{Department of Applied Physics and Stanford Institute for Materials and Energy Science, Stanford University, Stanford, California 94305, USA}
\affiliation{Japan Science and Technology Agency, Kawaguchi, 332-0012, Japan}
\date{\today}

\begin{abstract}
We report the violation of the Pauli limit due to intrinsic spin-orbit coupling in SrTiO$_3$ heterostructures. Via selective doping down to a few nanometers, a two-dimensional superconductor is formed, geometrically suppressing orbital pair-breaking. The spin-orbit scattering is exposed by the robust in-plane superconducting upper critical field, exceeding the Pauli limit by a factor of 4. Transport scattering times several orders of magnitude higher than for conventional thin film superconductors enables a new regime to be entered, where spin-orbit coupling effects arise non-perturbatively.
\end{abstract}
\maketitle

Unconventional superconductivity is a subject of great theoretical and experimental interest \cite{Sigrist_RMP1991,Fulde_PR1964,larkin_1964zz}. A central issue in this field is the discovery and understanding of non-trivial pairing mechanisms, such as the spin-triplet Cooper pair, which has been explicitly investigated in heavy fermions \cite{Sigrist_RMP1991}, Sr$_{2}$RuO$_{4}$ \cite{Mackenzie_RMP2003}, and crystals with broken inversion symmetry \cite{Bauer_PRL2004}. Recently, novel pairing has also been predicted in two-dimensional systems breaking inversion symmetry \cite{Gorkov_PRL2001,yip_PRB2002}. Experimentally, measurements of the superconducting upper critical field $H_{\mathrm{c2}}$ give vital information. In particular, violations of the Pauli paramagnetic limit \cite{Chandrasekhar_APL1962,Clogston_PRL1962} can be used to unravel the nature of the electron spins in the superconducting state. Notably, the presence of spin-orbit coupling (SOC) can be quantified \cite{Werthamer_PR1966}, as demonstrated by the $H_{\mathrm{c2}}$ studies of metal thin-film superconductors \cite{Tedrow_PRB1973} and bilayer systems where interface SOC drastically enhances $H_{\mathrm{c2}}$ \cite{wu_PRL2006}. 

Electron-doped SrTiO$_{3}$ (STO) has attracted much attention as the lowest-density superconductor \cite{koonce_pr1967} with high-mobility \cite{Tufte_PR1967}. These characteristics enable the creation of novel low dimensional systems \cite{Kozuka_NAT2009}, and are vital to shed light on the rich physics present at the LaAlO$_{3}$/SrTiO$_{3}$ (LAO/STO) interface, where the presence of the Rashba spin-orbit interaction has been discussed, affecting both the normal and superconducting state transport properties \cite{Caviglia_PRL2010,Shalom_PRL2010}. However, despite the fact that the conduction band structure of STO is similar to $p$-type GaAs \cite{mattheis_PRB1972a,Grbic_PRB2008}, the latter a model system for spintronics, the role of possible intrinsic SOC in the transport properties of doped STO is still unclear. 

\begin{figure}[b]
\includegraphics{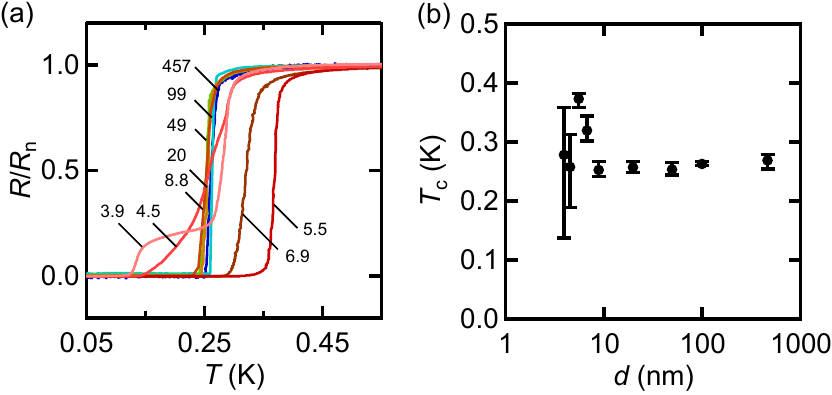}
\caption{\label{fig1}(color online) (a) Sheet resistance $R$, normalized by the normal-state value $R_{\mathrm{n}}$ as a function of temperature $T$. Numbers refer to the $\delta$-doped layer thickness $d$ in nm. (b) Superconducting transition temperature $T_{\mathrm{c}}$ versus $d$. $T_{\mathrm{c}}$ is defined by the temperature at the half value of the normal-state resistance; 10 \%-90 \% width of resistance is shown as an error bar.}
\end{figure}

\begin{figure*}[t]
\includegraphics{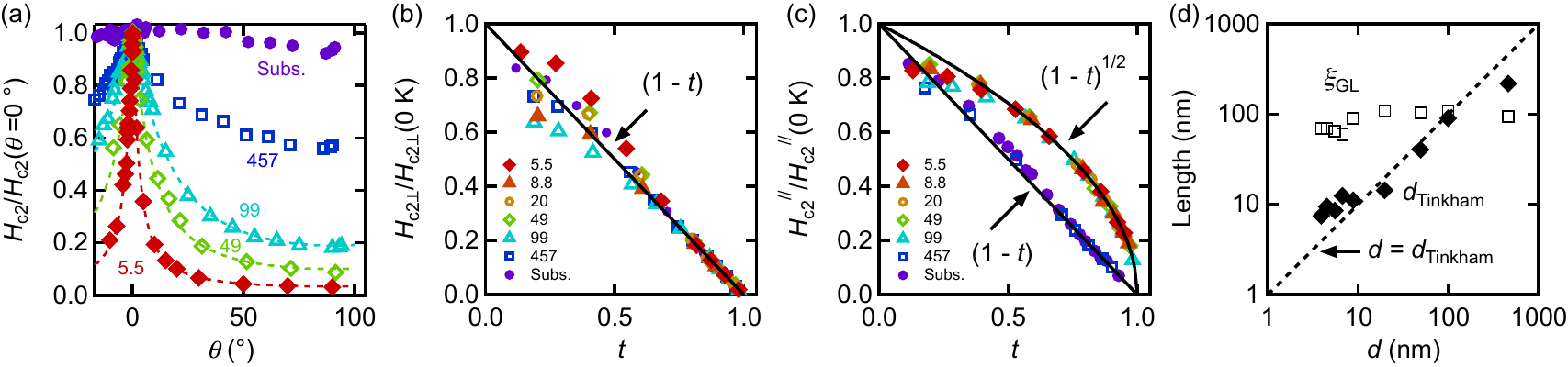}
\caption{\label{fig2}(color online) (a) Angular dependence of the upper critical field $H_{\mathrm{c2}}$ at 50 mK, normalized by the value at $\theta$ = 0 $^{\circ}$. Dashed curves are fitting results obtained to Tinkham's model. Results for representative samples (numbers refer to $d$ in nm) and a bulk 1 at$.$ \% NSTO substrate (Subs.) are shown. (b) Normalized perpendicular upper critical field $H_{\mathrm{c2}}^{\mathrm{\perp}}/H_{\mathrm{c2}}^{\mathrm{\perp}}$(0 K) plotted as a function of the reduced temperature $t=T/T_{\mathrm{c}}$. $H_{\mathrm{c2}}^{\mathrm{\perp}}$(0 K) was obtained by extrapolation to $T$ = 0 K from a fitting to the data in (b) over the range of $0.7 \le t \le 1$. (c) Normalized parallel upper critical field with same fitting procedure. (d) Ginzburg-Landau coherence length $\xi_{\mathrm{GL}}$ (open rectangles) and $d_{\mathrm{Tinkham}}$ (closed diamonds) versus $d$. Dashed line is $d$ = $d_{\mathrm{Tinkham}}$.}
\end{figure*}

In this Letter, we study the violation of the superconducting Pauli limit due to intrinsic SOC in a systematic series of symmetric, doped STO heterostructures. Using the $\delta$-doping technique, we selectively add Nb dopants in a narrow region inside an otherwise continuous undoped STO host crystal. As the thickness of the dopant layer is reduced, the destruction of superconductivity by orbital pair-breaking is geometrically suppressed, and the superconducting $H_{\mathrm{c2}}$ is enhanced for magnetic fields applied parallel to the dopant plane. In the thin regime, when the dopant layer is just a few nanometers thick, the superconductivity is robust beyond the conventional Pauli limit, demonstrating the presence of spin-orbit scattering (SOS) in the STO. Moreover, due to the absence of a surface or interface close to the dopant plane, and the spreading of the electron wavefunctions into the undoped STO, the electronic mean free path does not collapse as the dopant thickness decreases. Thus we preserve transport scattering times several orders of magnitude higher than for conventional thin film superconductors. In this regime, the intrinsic band SOC effects arise as a non-perturbative correction to the transport, despite the relatively long absolute SOS times.

The samples were fabricated with various thicknesses of 1 at$.$ \% doped Nb:SrTiO$_{3}$ (NSTO) films embedded between cap and buffer layers of undoped STO, using pulsed layer deposition. High-temperature growth, above $\sim$ 1050 $^{\circ}$C, in a low oxygen partial pressure of less than $10^{-7}$ Torr was chosen to achieve high-quality STO films, by managing the defect chemistry of the strontium and oxygen vacancies \cite{Kozuka_APL2010}. On a TiO$_{2}$ terminated STO (100) substrate, a 100 nm undoped STO buffer layer was first grown, followed by the 1 at$.$ \% NSTO layer with various thicknesses in the range 3.9 nm $\le d \le$ 457 nm. A 100 nm undoped STO cap layer was grown above the doped layer, to prevent surface depletion \cite{ohtomo_apl2004}. Post-annealing in a moderate oxidizing condition was used to fill oxygen vacancies formed during growth. Transport measurements were made using a standard four-probe method with sample cooling achieved using a dilution refrigerator with an {\it in-situ} rotator. For zero field measurements, the residual magnetic field was reduced below an absolute value of $\mu_0 H =  0.1$ mT, where $\mu_{\mathrm{0}}$ is the vacuum permeability.

All samples were superconducting at low temperatures, as shown in Fig$.$$\ \ref{fig1}$ (a). The transition temperatures $T_{\mathrm{c}}$ defined by the temperature below which the resistance was 50 \% of the normal-state value, were in the range 253 mK $\le T_{\mathrm{c}} \le$ 374 mK, as shown in Fig$.$$\ \ref{fig1}$ (b). All samples, except for the two thinnest, showed sharp 10 \%-90 \% transition widths ($\sim$ 10 mK). While samples with thickness $d$ $\geq$ 8.8 nm showed relatively constant $T_{\mathrm{c}}$ ($\sim$ 260 mK), several thinner samples showed a higher $T_{\mathrm{c}}$ while maintaining a sharp transition, suggesting possible changes to the superconducting properties close to the two-dimensional (2D) limit. We note that although the transition broadening in some of the thinner samples may relate to inhomogeneities, this is also reminiscent of the suggested Bose metal phase between the superconducting and insulating states \cite{Mason_PRL1999}. 

\begin{figure}[b]
\includegraphics{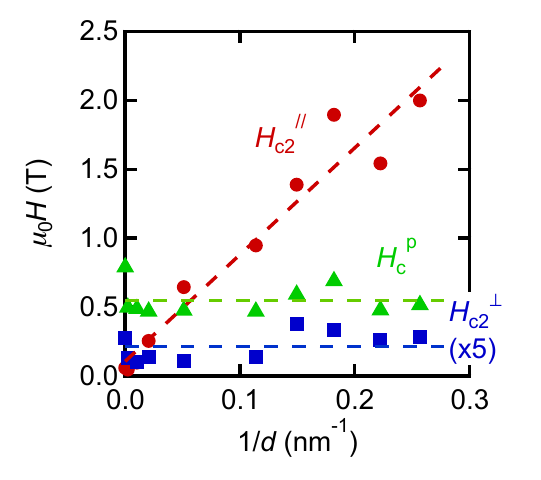}
\caption{\label{fig3}(color online) $H_{\mathrm{c2}}^{\mathrm{\parallel}}$ (circles), $H_{\mathrm{c2}}^{\mathrm{\perp}}$ (squares, scaled by factor of five), and the Pauli paramagnetic limit $H_{\mathrm{c}}^{\mathrm{p}}$ (triangles) plotted by $1/d$. $H_{\mathrm{c2}}^{\mathrm{\parallel}}$ and $H_{\mathrm{c2}}^{\mathrm{\perp}}$ are 50 mK data. Dashed lines are guides to the eye. }
\end{figure}

Firstly, the anisotropy of $H_{\mathrm{c2}}$ was used to measure the dimensionality of the superconductivity. We investigated the variation of $H_{\mathrm{c2}}$ by rotating the sample with respect to the magnetic field, as shown in Fig$.$$\ \ref{fig2}$ (a). A bulk 1 at$.$ \% NSTO substrate was also measured as a reference. As $d$ decreased, a clear modulation of $H_{\mathrm{c2}}$ as a function of the angle $\theta$ between the magnetic field and the sample plane was found. Here $H_{\mathrm{c2}}$ was defined as the field at which the resistance was half that of the normal state. For samples with $d$ $\leq$ 99 nm, excellent fits to these data could be made using Tinkham's model \cite{Tinkham_PR1963}, which is valid when the superconducting thickness is less than the Ginzburg-Landau coherence length $d_{\mathrm{Tinkham}} < \xi_{\mathrm{GL}}(0)$ \cite{supplementary}. These fits are shown in Fig$.$$\ \ref{fig2}$ (a).

\begin{figure*}[t]
\includegraphics{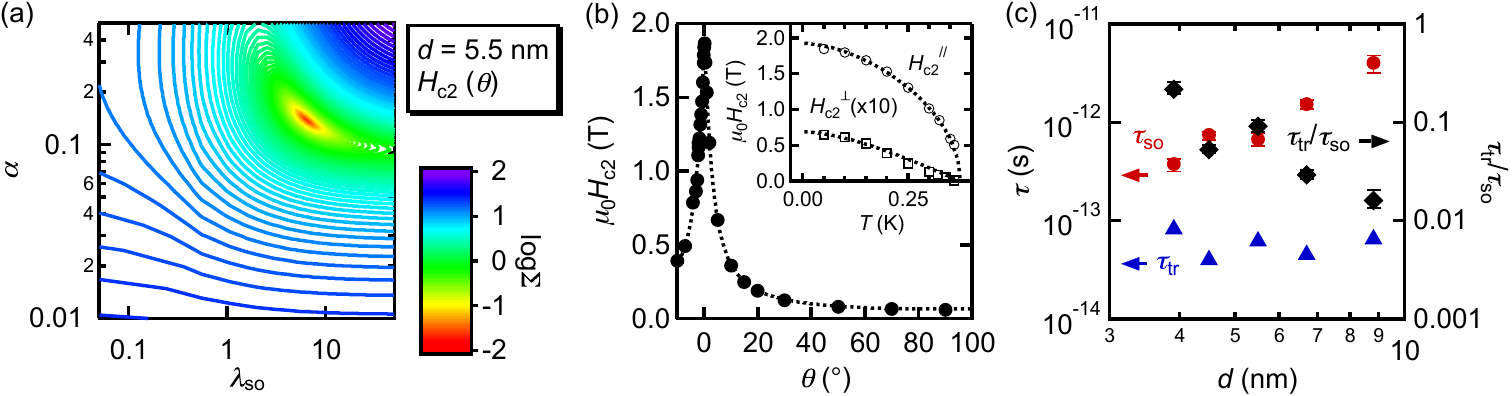}
\caption{\label{fig4}(color online) (a) Contour plot showing the deviation between the WHH simulation and the experiment by using $H_{\mathrm{c2}}(\theta)$ data for $d$ = 5.5 nm. The minimum point is located at $\lambda$ = 6.4, $\alpha$ = 0.14. (b) $H_{\mathrm{c2}}(\theta)$ of 5.5 nm thick sample at 50 mK. Dotted line is the best fit obtained from the WHH simulation. Inset: $H_{\mathrm{c2}}^{\mathrm{\perp}}(T)$ and $H_{\mathrm{c2}}^{\mathrm{\parallel}}(T)$ data and the WHH theory fit (dotted line). (c) Variation of $\tau_{\mathrm{so}}$ and $\tau_{\mathrm{tr}}$ with $d$ for the five thinnest samples, obtained from best fits to the WHH theory. An error bar of $\tau_{\mathrm{so}}$ is given by assuming 10 \% thickness variation of the superconducting layer. The ratio $\tau_{\mathrm{tr}}/\tau_{\mathrm{so}}$ is also shown on the right axis.}
\end{figure*}

The dimensional crossover of superconductivity is more clearly demonstrated by the temperature dependence of $H_{\mathrm{c2}}$, therefore we next measured $H_{\mathrm{c2}}^{\mathrm{\perp}}(t)$ and $H_{\mathrm{c2}}^{\mathrm{\parallel}}(t)$, the out-of-plane ($\theta$ = 90 $^{\circ}$) and in-plane ($\theta$ = 0 $^{\circ}$) upper critical fields respectively (here $t = T/T_c$), as shown in Figs$.$$\ \ref{fig2}$ (b) and (c). In the perpendicular field geometry, all samples showed a linear temperature dependence. In the parallel field geometry, for $d$ $\leq$ 99 nm, however, $H_{\mathrm{c2}}^{\mathrm{\parallel}}(t)$ showed a clear square root form, which is characteristic of the 2D superconducting state. These data clearly demonstrate a three-dimensional (3D) to 2D crossover of the superconducting character as a function of $d$. By estimating $d_{\mathrm{Tinkham}}$ and $\xi_{\mathrm{GL}}(0)$ from the $H_{\mathrm{c2}}^{\mathrm{\perp}}$ and $H_{\mathrm{c2}}^{\mathrm{\parallel}}$ data, we find that $d_{\mathrm{Tinkham}}$ decreases in proportion to the growth thickness $d$, and in the thinnest sample is much smaller than $\xi_{\mathrm{GL}}$(0) $\approx$ 100 nm, as plotted in Fig$.$$\ \ref{fig2}$ (d), confirming the 2D nature of the superconductivity. 

A crucial and intriguing aspect of the $H_{\mathrm{c2}}^{\mathrm{\parallel}}(t)$ data is the violation of the Pauli paramagnetic limit. The Pauli paramagnetic limiting field \cite{Chandrasekhar_APL1962,Clogston_PRL1962} is given by $H_{\mathrm{c}}^{\mathrm{p}} =  \Delta_\mathrm{0} / \sqrt{2}\mu_{\mathrm{B}}$, where $\mu_{\mathrm{B}}$ is the Bohr magneton (with a $g$-factor of two), and $\Delta_\mathrm{0} = 1.76k_{\mathrm{B}}T_{\mathrm{c}}$ is the BCS superconducting gap for a weak-coupling superconductor, where $k_{\mathrm{B}}$ is Boltzmann's constant. This limit is appropriate, since via tunneling bulk doped STO is known to be in the weak-coupling regime \cite{Binnig_SSC1974}. The variation of $H_{\mathrm{c2}}^{\mathrm{\perp}}$, $H_{\mathrm{c2}}^{\mathrm{\parallel}}$, and $H_{\mathrm{c}}^{\mathrm{p}}$ as a function of $1/d$ is shown in Fig$.$$\ \ref{fig3}$. $H_{\mathrm{c2}}^{\mathrm{\parallel}}$ exceeds the Pauli limiting field $H_{\mathrm{c}}^{\mathrm{p}}$ by a factor of more than four in the thinnest sample, while $H_{\mathrm{c2}}^{\mathrm{\perp}}$ and $H_{\mathrm{c}}^{\mathrm{p}}$ remained essentially constant. 

In the case of a 2D superconductor in a parallel magnetic field, if the sample is thin enough that orbital depairing is suppressed, spin paramagnetism is the dominant mechanism for destroying superconductivity \cite{Tedrow_PRB1973}. In the presence of SOS, however, $H_{\mathrm{c2}}^{\mathrm{\parallel}}$ can be robust beyond the Pauli limit. It should be noted that the renormalization of normal-state properties by many-body effects \cite{Orlando_PRB1979} can also enhance the Pauli limit. However, Shubnikov-de Haas oscillations in these $\delta$-doped samples showed that the electron mass is consistent with the band structure at low temperature \cite{Kozuka_NAT2009}. Given the low electron-density, strong correlation effects near half-filling are also absent.

To investigate the effects of SOS in more detail, we performed a numerical fit of the $H_{\mathrm{c2}}(\theta,t)$ data to the Werthamer-Helfand-Hohenberg (WHH) theory \cite{Werthamer_PR1966} taking into account corrections for the thin film case \cite{supplementary}. Within this theory, the two crucial fitting parameters, the orbital depairing parameter $\alpha$ and the SOS rate $\lambda_{\mathrm{so}}$, are given by
\begin{equation}
\alpha = \frac{\hbar}{2m^{*}D},
\end{equation}
\begin{equation}
\lambda_{\mathrm{so}} = \frac{2\hbar}{3\pi k_{\mathrm{B}}T_{\mathrm{c}}\tau_{\mathrm{so}}},
\end{equation}
where $\hbar$ is the Plank constant divided by 2$\pi$, $m^{*}$ is the effective electron mass, $D=v_{\mathrm{F}}^{2}\tau_{\mathrm{tr}}/3$ is the diffusion constant, and $v_{\mathrm{F}}$ is the Fermi velocity. $\tau_{\mathrm{tr}}$ and $\tau_{\mathrm{so}}$ are, respectively, the transport and SOS times. For various $\alpha$ and $\lambda_{\mathrm{so}}$, we calculated the sum of the squares of the differences between the WHH model and the $H_{\mathrm{c2}}(\theta)$ data (we denote this sum as $\Sigma$), for the case $d= 5.5$ nm, as shown in Fig$.$$\ \ref{fig4}$ (a). A unique minimum value of $\Sigma$ was found, giving an excellent fit to the experimental data, as shown in Fig$.$$\ \ref{fig4}$ (b). We obtain $\tau_{\mathrm{tr}}$ = $6.2\times 10^{-14}$ s and $\tau_{\mathrm{so}}$ = $6.8\times 10^{-13}$ s, from $\alpha$ and $\lambda_{\mathrm{so}}$, respectively. This clearly indicates that the SOS is a highly significant factor affecting the normal and superconducting transport properties, in spite of its relatively small absolute value. This is due to an appealing point of the $\delta$-doped STO structures, where $\tau_{\mathrm{tr}}$ is several orders of magnitudes higher than for conventional metal superconductors \cite{Hake_APL1967}.

By performing similar analysis on data for the other samples where $H_{\mathrm{c2}}^{\mathrm{\parallel}}$ exceeded $H_{\mathrm{c}}^{\mathrm{p}}$, we obtain the thickness dependences of $\tau_{\mathrm{so}}$ and $\tau_{\mathrm{tr}}$, as shown in Fig$.$$\ \ref{fig4}$ (c). A clear decrease of $\tau_{\mathrm{so}}$ with decreasing $d$ is found, while $\tau_{\mathrm{tr}}$ is relatively unchanged. This dependence suggests that the SOS is not dominated by either the Elliott-Yafet mechanism where $\tau_{\mathrm{so}} \propto \tau_{\mathrm{tr}}$, or the D'yakonov-Perel' mechanism where $\tau_{\mathrm{so}} \propto 1 / \tau_{\mathrm{tr}}$ \cite{Zutic_RMP2004}. This rather unexpected result suggests that the SOS observed has a different origin. We next clarify this point by comparison with other systems.

We emphasize that this combination of SOC with high mobility conduction electrons places our system is in a different regime compared to other thin film structures that violate the Pauli limit. For example, the use of heavy atoms to induce SOC in superconducting bilayers has been studied \cite{wu_PRL2006}. However, in this case, as is usual for conventional superconducting thin films, the mean free path collapses in the thin limit. A similar collapse occurs with substrate gating at the LAO/STO heterointerface \cite{Bell_PRL2009} where an asymmetric confining potential and Rashba SOC is expected \cite{Caviglia_PRL2010,Shalom_PRL2010}. $\delta$-doping is a crucial determinant for this difference: since there is no obvious surface or interface surrounding the conducting layer, the scattering length due to disorder is unchanged (and even increased) with decreasing $d$ \cite{Kozuka_APL2010_2}. Additionally, the symmetry of the structure, giving rise to zero net effective electric field, means that Rashba SOC is absent. 

We can thus interpret this as intrinsic SOC of STO, due to the $d$-orbitals of the Ti atoms \cite{mattheis_PRB1972a}. Indeed the bulk conduction bands of STO have a similar structure to the valence bands of GaAs, where non-perturbatively large SOC has been demonstrated \cite{Grbic_PRB2008}. However in the case of STO there are few studies of electron SOS. It should be noted that calculations indicate that these confined $\delta$-doped samples have a multiple subband structure \cite{Kozuka_NAT2009}, therefore the change of $\tau_{\mathrm{so}}$ observed may relate to intersubband-induced spin-orbit interaction \cite{Bernardes_PRL2007}, or intersubband scattering \cite{Hansen_PRB1993}, which in turn are influenced by changes of the band structure as $d$ changes.

The fact that the energy scale of the observed SOC ($\sim$2 meV) is bigger than the superconducting gap ($\sim$40 $\mu$eV) suggests the possibility of mixed spin-triplet and singlet states, giving rise to novel superconducting states in these low dimensional layers \cite{GINZBURG_PR1964,Gorkov_PRL2001,yip_PRB2002}. Moreover, in the normal state, the combination of high mobility conduction and SOC imply that STO can be usefully employed in a wide range of controllable spintronic architectures, including the spin Hall effect, which have until now been dominated by metals and traditional semiconductors. Thus this system is ideal for future measurements, which can be made over a range of densities both via growth control \cite{Kim_condmat2011}, and field-effect gating, the latter simultaneously introducing Rashba contributions to the SOC \cite{Koga_prl2002}.

The authors thank M. Lippmaa for experimental assistance, M$.$ R$.$ Beasley, A$.$ Bhattacharya, A$.$ Kapitulnik, K$.$ H$.$ Kim, A$.$ H$.$ MacDonald, A$.$ F$.$ Morpurgo, B$.$ Spivak, and H$.$ Takagi for discussions. C$.$B$.$ and H$.$Y$.$H$.$ acknowledge support by the Department of Energy, Office of Basic Energy Sciences, Division of Materials Sciences and Engineering, under contract DE-AC02-76SF00515. M$.$K$.$ acknowledges support from the Japanese Government Scholarship Program of the Ministry of Education, Culture, Sport, Science and Technology, Japan.

\subsection*{Supplemental Material - Intrinsic Spin-Orbit Coupling in Superconducting $\delta$-doped SrTiO$_3$ Heterostructures}
\setcounter{figure}{0}
\setcounter{equation}{0}
\renewcommand{\thefigure}{S\arabic{figure}}

\subsection*{Tinkham's model}
The magnetic-field response of a two-dimensional superconductor can be described by Tinkham's model \cite{Tinkham_PR1963s}, which assumes that the thickness of the superconductor is thinner than the Ginzburg-Landau coherence length, $d_{\mathrm{Tinkham}} < \xi_{\mathrm{GL}}$. It should be noted that the model does not include the effects of spin-orbit scattering or the Pauli paramagnetic limit, and assumes an isotropic superconducting wavefunction. According to the model, the angular dependence of the upper critical field can be shown to be
	\begin{equation}
	\left| \frac{H_{\mathrm{c2}}(\theta)\sin\theta}{H_{\mathrm{c2}}^{\mathrm{\perp}}}\right|+\left( \frac{H_{\mathrm{c2}}(\theta)\cos\theta}{H_{\mathrm{c2}}^{\mathrm{\parallel}}} \right)^{2}=1,\label{eq:thetadep_Hc}
	\end{equation}
	where $\theta$ is the angle between the magnetic field and the sample plane. The temperature dependence of the upper critical field in perpendicular and parallel field geometry is given by
\begin{equation}
	H_{\mathrm{c2}}^{\mathrm{\perp}}(t) = \frac{\Phi_{0}}{2\pi\xi_{\mathrm{GL}}(0)^{2}}\left( 1-t \right), \label{eq:Tdep_Hcperp}
	\end{equation}
	\begin{equation}		
	H_{\mathrm{c2}}^{\mathrm{\parallel}}(t) = \frac{\Phi_{0}\sqrt{12}}{2\pi\xi_{\mathrm{GL}}(0)d_{\mathrm{Tinkham}}}\left( 1-t \right)^{\frac{1}{2}}, \label{eq:Tdep_Hcparall}
	\end{equation}		
	where $t ={T}/{T_{\mathrm{c}}}$ is the reduced temperature, $\Phi_0 = h/2e = 2.07 \times 10^{-15} \;\mathrm{Wb}$ is the flux quantum, and $\xi_{\mathrm{GL}}$(0) is the Ginzburg-Landau coherence length extrapolated to $T$ = 0 K. From Eqs$.$\ \ref{eq:Tdep_Hcperp} and \ref{eq:Tdep_Hcparall}, $d_{\mathrm{Tinkham}}$ and $\xi_{\mathrm{GL}}(0)$ can be found using
	\begin{equation}		
	d_{\mathrm{Tinkham}} = \sqrt{\frac{6\Phi_{0}H_{\mathrm{c2}}^{\mathrm{\perp}}}{\pi(H_{\mathrm{c2}}^{\mathrm{\parallel}})^{2}}}, \label{eq:dtinkham}
	\end{equation}
	\begin{equation}		
	\xi_{\mathrm{GL}}(0) = \sqrt{\frac{\Phi_{0}}{2\pi H_{\mathrm{c2}}^{\mathrm{\perp}}}}. \label{eq:xi_gl}
	\end{equation} Thus, $d_{\mathrm{Tinkham}}$ can be calculated by measurement of $H_{\mathrm{c2}}^{\mathrm{\perp}}$ and $H_{\mathrm{c2}}^{\mathrm{\parallel}}$ of a sample experimentally. As noted by Ben Shalom $et$ $al.$ \cite{Shalom_PRL2010s}, in the case of the LaAlO$_3$/SrTiO$_3$ interface, the value of $d_{\mathrm{Tinkham}}$ is an upper bound on the thickness. In our case we find good agreement between the grown dopant layer thickness $d$ and $d_{\mathrm{Tinkham}}$ in thick samples, but $d_{\mathrm{Tinkham}}$ deviates slightly from $d$ as $H_{\mathrm{c2}}^{\mathrm{\parallel}}$ exceeds the Pauli limiting field in the thinnest samples ($d \leq 8.8$ nm), indicating a limit of the model.

\subsection*{WHH theory}

The Werthamer-Helfand-Hohenberg (WHH) theory \cite{Werthamer_PR1966s} was used to more quantitatively fit the superconducting upper critical field, $H_{\mathrm{c2}}$, data in the main text, in order to determine the spin-orbit scattering time in the system. Within this theory $H_{\mathrm{c2}}$ is the implicit solution of the equation
\begin{eqnarray}
\nonumber &\mathrm{ln}t+\left(\frac{1}{2}+\frac{i\lambda_{\mathrm{so}}}{4\gamma}\right)\psi\left(\frac{1}{2}+\frac{\bar{h}+\frac{1}{2}\lambda_{\mathrm{so}}+i\gamma}{2t}\right)\\
\nonumber &+\left(\frac{1}{2}-\frac{i\lambda_{\mathrm{so}}}{4\gamma}\right)\psi\left(\frac{1}{2}+\frac{\bar{h}+\frac{1}{2}\lambda_{\mathrm{so}}-i\gamma}{2t}\right)-\psi\left(\frac{1}{2}\right)=0, \label{eq:WHH}
\end{eqnarray}

where $\psi$ is the digamma function. With a slight correction \cite{fetter_parksbook} from the original WHH paper, the terms are defined as
\begin{equation}
\bar{h} = \frac{DeH_{\mathrm{c2}}}{\pi k_{\mathrm{B}}T_{\mathrm{c}}},
\end{equation}
\begin{equation}
\lambda_{\mathrm{so}} = \frac{2\hbar}{3\pi k_{\mathrm{B}}T_{\mathrm{c}}\tau_{\mathrm{so}}},
\end{equation}
\begin{equation}
\gamma = \sqrt{\mathstrut(\alpha\bar{h})^{2}-\frac{1}{4}\lambda_{\mathrm{so}}^{2}},
\end{equation}
\begin{equation}
\alpha = \frac{\hbar}{2mD},
\end{equation}
where $D$ is the diffusion constant, $\tau_{\mathrm{so}}$ the spin-orbit scattering time, and $m$ the electron mass. To include the effect of the finite thickness of a thin film, the term $\bar{h}$ in Eq$.$$\ \ref{eq:WHH}$ should be replaced by $\bar{h}^{\mathrm{ang}}(\theta)$, which is given by \cite{Aoi_PRB1973}
\begin{equation}\label{aoi_model}
\bar{h}^{\mathrm{ang}}(\theta) = \frac{D}{2\pi k_{\mathrm{B}}T_{\mathrm{c}}}\left(2eH_{\mathrm{c2}}|\sin(\theta)|+\frac{1}{3\hbar}(deH_{\mathrm{c2}}\cos(\theta))^{2}\right),
\end{equation}
where $d$ is thickness of superconducting layer. 

In the case of $\alpha = \lambda_{\mathrm{so}} = 0$ (no spin-orbit coupling, but also no Pauli paramagnetic limit), the above formula is reduced to the orbital term only, where $H_{\mathrm{c2}}$ is the solution of
\begin{equation}\label{eq:orbitalonly}
\mathrm{ln}t+\psi\left(\frac{1}{2}+\frac{\bar{h}}{2t}\right)-\psi\left(\frac{1}{2}\right)=0.
\end{equation}
In fitting the data, the orbital only case (Eq. \ref{eq:orbitalonly}) could not accurately fit the data for samples with $H_{\mathrm{c2}}^{\mathrm{\parallel}}$ larger than the Pauli paramagnetic limit. However we could obtain a very successful fit with the full WHH theory (Eq. \ref{eq:WHH}), as discussed and shown in the main text (Fig. 4). 

\begin{figure}[b]
\includegraphics{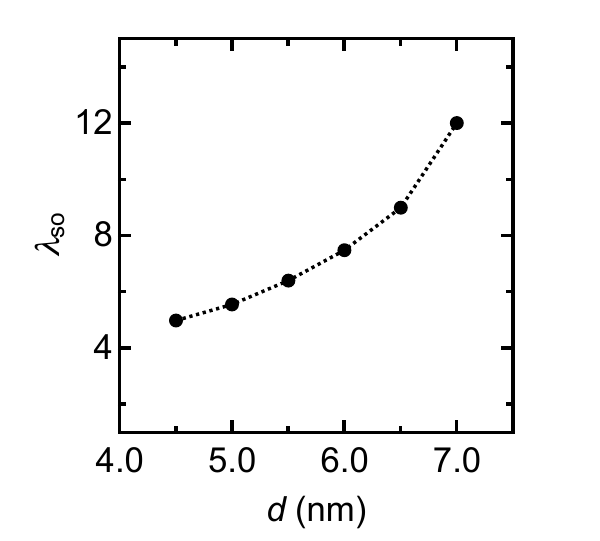}
\caption{Optimal value of $\lambda_{\mathrm{so}}$ from the WHH fitting for the $d=5.5$ nm sample, depending on the thickness $d$ used in the model.}
\label{figS1}
\end{figure}

Several sources of error should be considered in this fit. Firstly, it has been argued that cooling superconducting ultra-thin films below 60 mK is extremely difficult \cite{parendo_prb2006}, thus the $H_{\mathrm{c2}}(t)$ data may artificially saturate at low temperatures due to a lack of cooling. However, by fitting to data at temperatures only above 75 mK, we find an increase of the value of $\tau_{\mathrm{so}}$ of only 2 \% compared to the full fitting curve. Secondly, the influence of error in the value of $d$ used in Eq$.$ \ref{aoi_model} can also be considered. This effect is demonstrated in Fig$.$ $\ref{figS1}$, where the value of $\tau_{\mathrm{so}}$ obtained from the fitting is plotted against $d$ for the $d$ = 5.5 nm sample. As is clear, sensitivity of the fit to variation in $d$ gives rise to variation in $\tau_{\mathrm{so}}$; we obtain $\tau_{\mathrm{so}} = (6.8 \pm 1.0) \times 10^{-13}$ s for $d = 5.5 \pm 0.5$ nm.

\begin{figure}[t]
\includegraphics{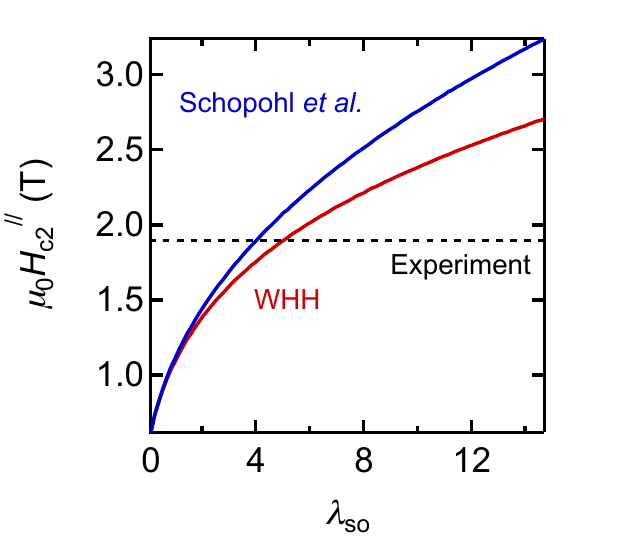}
\caption{Variation of $H_{\mathrm{c2}}^{\mathrm{\parallel}}$ with $\lambda_{so}$ using a non-perturbative spin-orbit coupling model, compared to the original WHH model, adapted from Schopohl $et$ $al.$ \cite{Schopohl_PBC1981}.}
\label{figS2}
\end{figure}

We note that the WHH theory makes various simplifying assumptions: the superconductor should be in the dirty limit, where the electron mean free path is shorter than the BCS coherence length $\ell \ll \xi_{\mathrm{BCS}}$. Secondly, the spin-orbit scattering time is less than the total scattering time $\tau_{\mathrm{so}} \ll \tau_{\mathrm{tr}}$. In order to estimate these parameters, we assume a single band approximation with spherical Fermi surface and used an electron effective mass $m^{*} = 1.24 m_{\mathrm{0}}$, where $m_{\mathrm{0}}$ is the bare electron mass, extracted from Shubnikov-de Haas oscillations \cite{Kozuka_Nat2009}. In the $d$ = 5.5 nm sample, we found $\ell \approx$ 100 nm, $\xi_{\mathrm{BCS}} \approx$ 470 nm, $\tau_{\mathrm{tr}} \approx 6.2 \times 10^{-14}$ s. Therefore, we conclude that our system is in the dirty limit, and spin-orbit coupling can be treated as a perturbation. This however becomes less clear with decreasing thickness, for which $\tau_{\mathrm{tr}}/\tau_{\mathrm{so}} \approx 0.1$. As a further check, we used a non-perturbative theory proposed by Schopohl $et$ $al.$ \cite{Schopohl_PBC1981}. The value of $H_{\mathrm{c2}}^{\mathrm{\parallel}}$ for various $\lambda_{so}$ using this theory is shown in Fig$.$ $\ref{figS2}$, along with the original WHH model (neglecting the finite size corrections of Aoi $et$ $al.$ \cite{Aoi_PRB1973}). Since in absolute terms the observed values of $H_{\mathrm{c2}}^{\mathrm{\parallel}}$ are not large due to the relatively low $T_{\mathrm{c}}$, we estimate an error of only $\sim$ 20 \% in the determination of $\tau_{\mathrm{so}}$ by using the original WHH model, as shown, which does not significantly affect the result of the original fitting.

\end{document}